\documentclass[jkps,preprint,fleqn,showpacs,showkeys]{revtex4}
\usepackage[pdftex]{graphicx}
\usepackage{amssymb}
\usepackage{amsmath}
\usepackage{bm}
\begin{document}
\setcounter{page}{0}
\title[]{Evaluation of Stopping Powers of Superheavy Ions with $\it{Z}$ up to 124 in Al and U}
\author{Yong Hee \surname{Chung}}
\email{yhchung@hallym.ac.kr}
\thanks{Fax: +82-33-256-3421}
\affiliation{Department of Chemistry, Hallym University, Chuncheon 200-702}

\date[]{}

\begin{abstract}

Electronic and nuclear stopping powers of superheavy ions with $\it{Z}$ up to 124 and $\it{A}$=300 in Al and U were estimated in the energy range of 0.01-0.20 MeV/u. The corresponding stopping powers were estimated using stopping powers of ions with 6$\leq$$\it{Z}$$\leq$92 obtained from SRIM. The results were compared with those deduced from Northcliffe and Schilling$'$s stopping-power tables and with experimental data for $^{16}$O and $^{19}$F ions in Al. Estimated electronic stopping powers of the light ions agreed better with their experimental values above 0.10 MeV/u than the SRIM data. 
\end{abstract}



\maketitle

\section{INTRODUCTION}

In the study of the excitation function of superheavy ions their energy losses have to be known along with their masses, which can be taken from one of the mass tables \cite{NPA96Myers, AD95Moller, NPA03Audi}. No energy loss data are available to heavy ions with $\it{Z}$$>$103. The electronic stopping-power and range tables for ions with 1$\leq$$\it{Z}$$\leq$103 at 38 energies ranging from 0.0125 to 12 MeV/u in 24 media have been reported by Northcliffe and Schilling \cite{NDT70Northcliffe}. Hubert  $\it{et}$ $\it{al.}$ \cite{AD90Hubert} reported electronic stopping-power and range tables for 2.5-500 MeV/u ions with 2$\leq$$\it{Z}$$\leq$103 in 36 solid materials. However, nuclear stopping powers were not listed in both tables. Electronic and nuclear stopping powers of ions with $\it{Z}$$\leq$92 in various media can be obtained from SRIM \cite{SRIM}, but stopping powers of charged particles with $\it{Z}$$>$92 are not available. There are scarce experimental data and expected values of stopping powers for heavy ions with $\it{Z}$$>$103, but the stopping power of $^{289}$Fl ($\it{Z}$=114) in Mylar was reported \cite{NIM10Wittwer}.

In the previous work \cite{IJMPE10Chung} electronic and nuclear stopping powers of a superheavy ion with $\it{Z}$=120 and $\it{A}$=300 at 26 energies ranging from 0.0177 to 0.141 MeV/u in UF$_4$, He, Mylar and butane have been estimated. Since hot fusion could be used in producing elements with $\it{Z}$$\geq$120, the energy range has been expanded to 0.20 MeV/u to estimate stopping powers of a heavy ion with $\it{Z}$=120 and $\it{A}$=300 in Al and UO$_2$ \cite{CJP14Chung}. In this work stopping powers of superheavy ions with $\it{Z}$$>$120 in Al and those with $\it{Z}$$\geq$120 in U are estimated at 35 energies ranging from 0.01 to 0.20 MeV/u. The estimation has been evaluated for light ions. 

\section{Energy loss of superheavy ions in media }

Energy loss of superheavy ions in media arises from their electronic and nuclear stopping powers. The electronic stopping powers were deduced by fitting data from SRIM with a function with six parameters \cite{IJMPE10Chung} or by fitting them with a following function \cite{CJP14Chung}:
\begin{equation} 
(-dE/dx)_e/Z^2 = a_1 +a_2 $exp$[-a_3/Z^{2/3}] 
\label{eq:one}
\end{equation} 
The nuclear stopping powers of the superheavy ion were deduced with a fifth-order polynomial fit with respect to $\it{Z}$ \cite{IJMPE10Chung,CJP14Chung}. Electronic stopping power of $^{289}$Fl in Mylar at 24.7$\pm$4.0 MeV deduced by Eq. \eqref{eq:one} was 30.2$\pm$2.7 MeV/(mg/cm$^2$) with its corresponding nuclear stopping powers of 8.3$\pm$1.2 MeV/(mg/cm$^2$) \cite{CJP14Chung}. The resulting total stopping power of 38.5$\pm$3.0 MeV/(mg/cm$^2$) agreed quite well with the experimental one of 34.9$\pm$15.4 MeV/(mg/cm$^2$), implying that this estimation is applicable to superheavy ions with $\it{Z}$=120 or even higher.

\subsection{Stopping powers of superheavy ions with $\it{Z}$$>$120 and $\it{A}$=300 in Al}

\begin{figure}
\includegraphics[width=10cm]{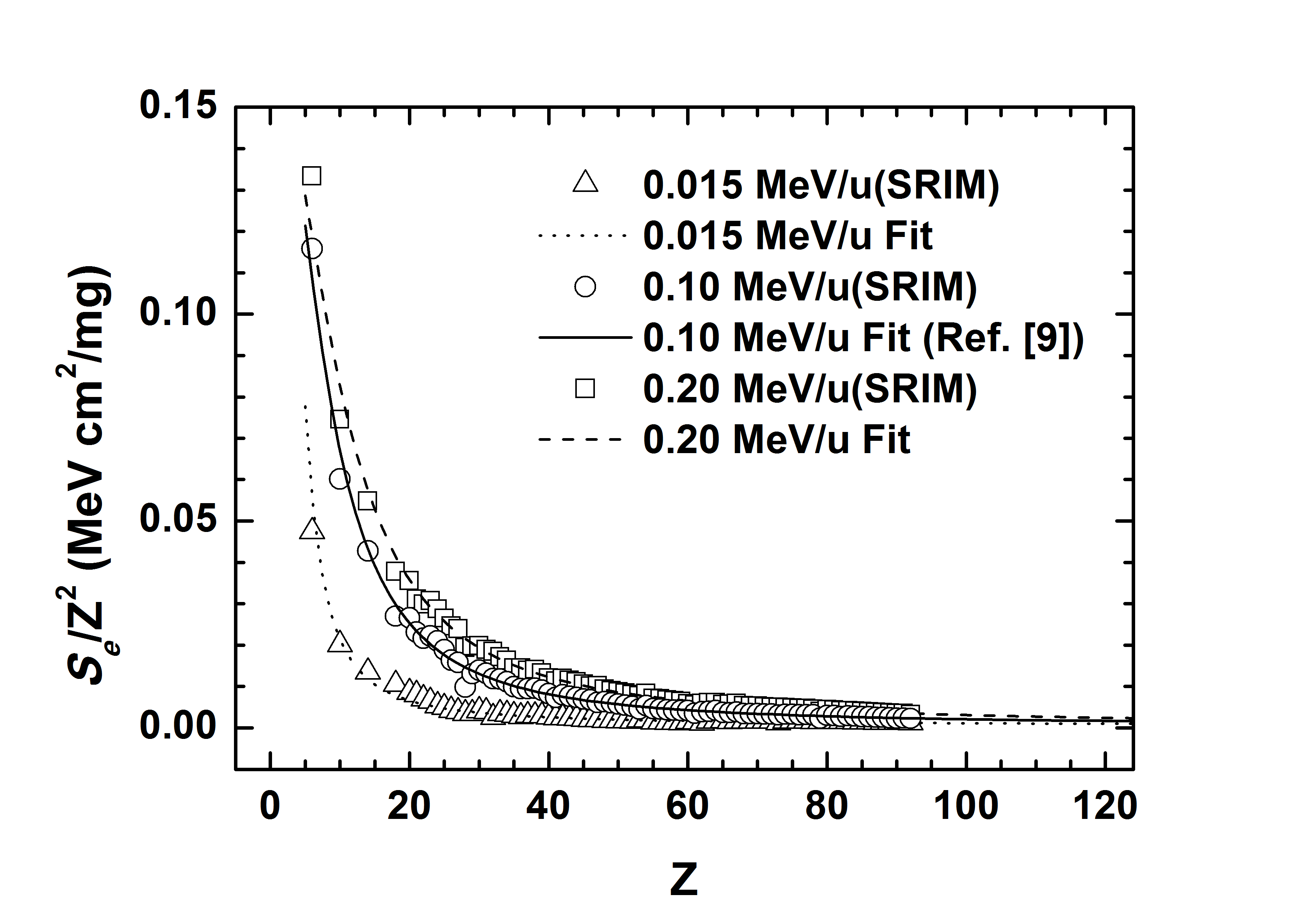}
\caption{Electronic stopping powers of the ions in Al divided by $\it{Z^2}$ at 0.015, 0.10 and 0.20 MeV/u. Open symbols are obtained from SRIM \cite{SRIM} and the lines refer to their corresponding fits with Eq. \eqref{eq:one}.}
\label{fig1}
\end{figure}

\begin{figure}
\includegraphics[width=10cm]{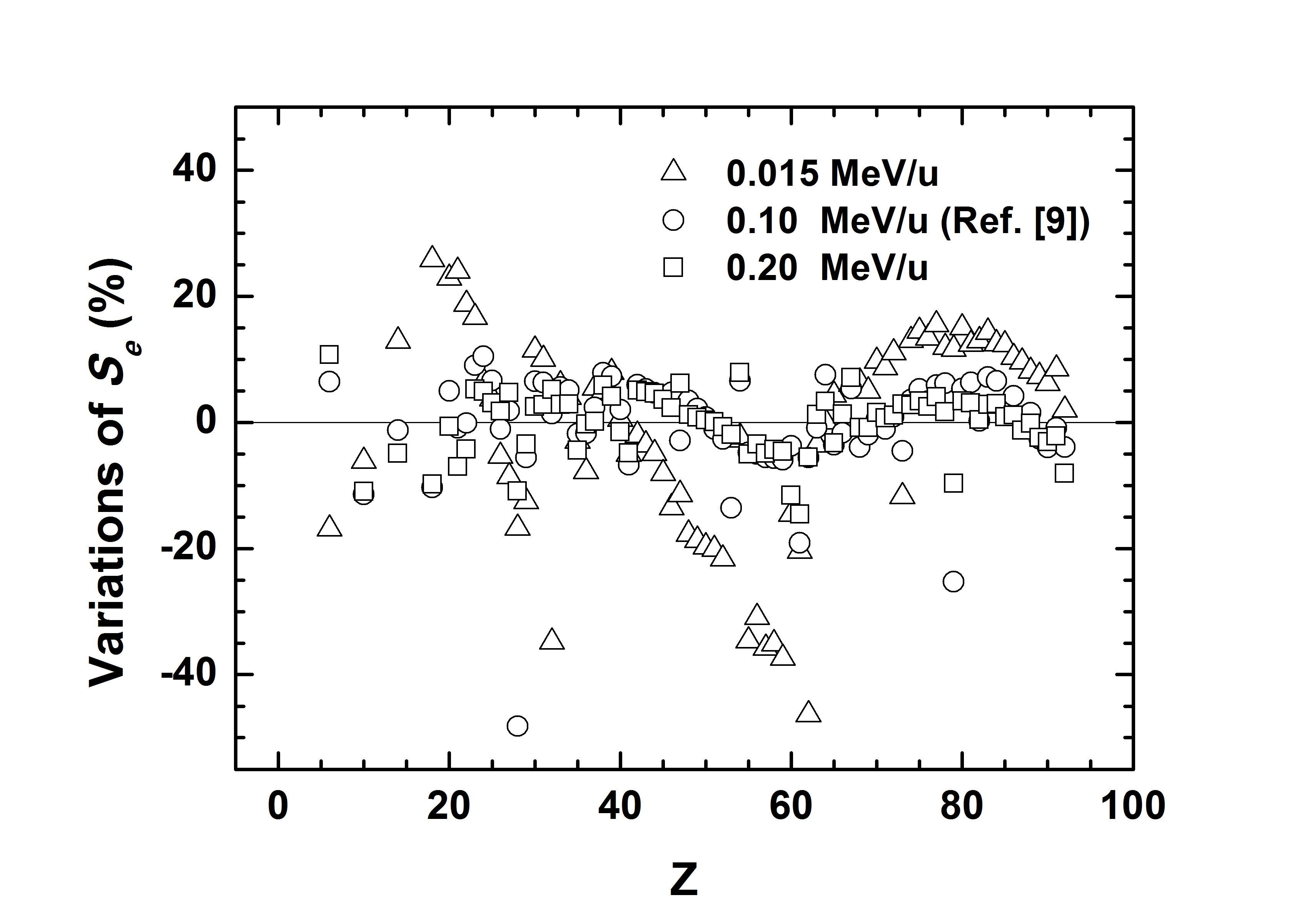}
\caption{Variations of the fits in Fig. \ref{fig1}.}
\label{fig2}
\end{figure}

Electronic and nuclear stopping powers of 77 ions ranging from C to U in Al at 35 energies have been obtained from SRIM \cite{SRIM}. Their atomic numbers are 6, 10, 14, 18, 20, 21, 22,..., 91, and 92. Their 35 energies range from 0.01 to 0.20 MeV/u. The electronic and nuclear stopping powers at each energy have been used to extrapolate the corresponding stopping powers of heavy ions with $\it{Z}$=120 and $\it{A}$=300 with Eq. \eqref{eq:one} and the fifth-order polynomial fit, respectively \cite{CJP14Chung} and to further extrapolate those of heavy ions with $\it{Z}$=122 and 124 and $\it{A}$=300. The fits and the resulting variations at 0.015, 0.10 and 0.20 MeV/u are shown in Figs. \ref{fig1} and \ref{fig2}, respectively. They show that the fits work well with most of the ions except five ions with variations larger than 35 \% at  0.015 MeV/u and three ions with variations larger than 15 \% at 0.10 MeV/u. The ions at 0.015 MeV/u are Ge with 35 \%, Cs with 35 \%, Ce with 36 \%, Pr with 37 \% and Sm with 46 \% variation, while those at 0.10 MeV/u are Ni with 48 \%, Pm with 19 \% and Au with 25 \% variation \cite{CJP14Chung}. 

\begin{figure}
\includegraphics[width=10cm]{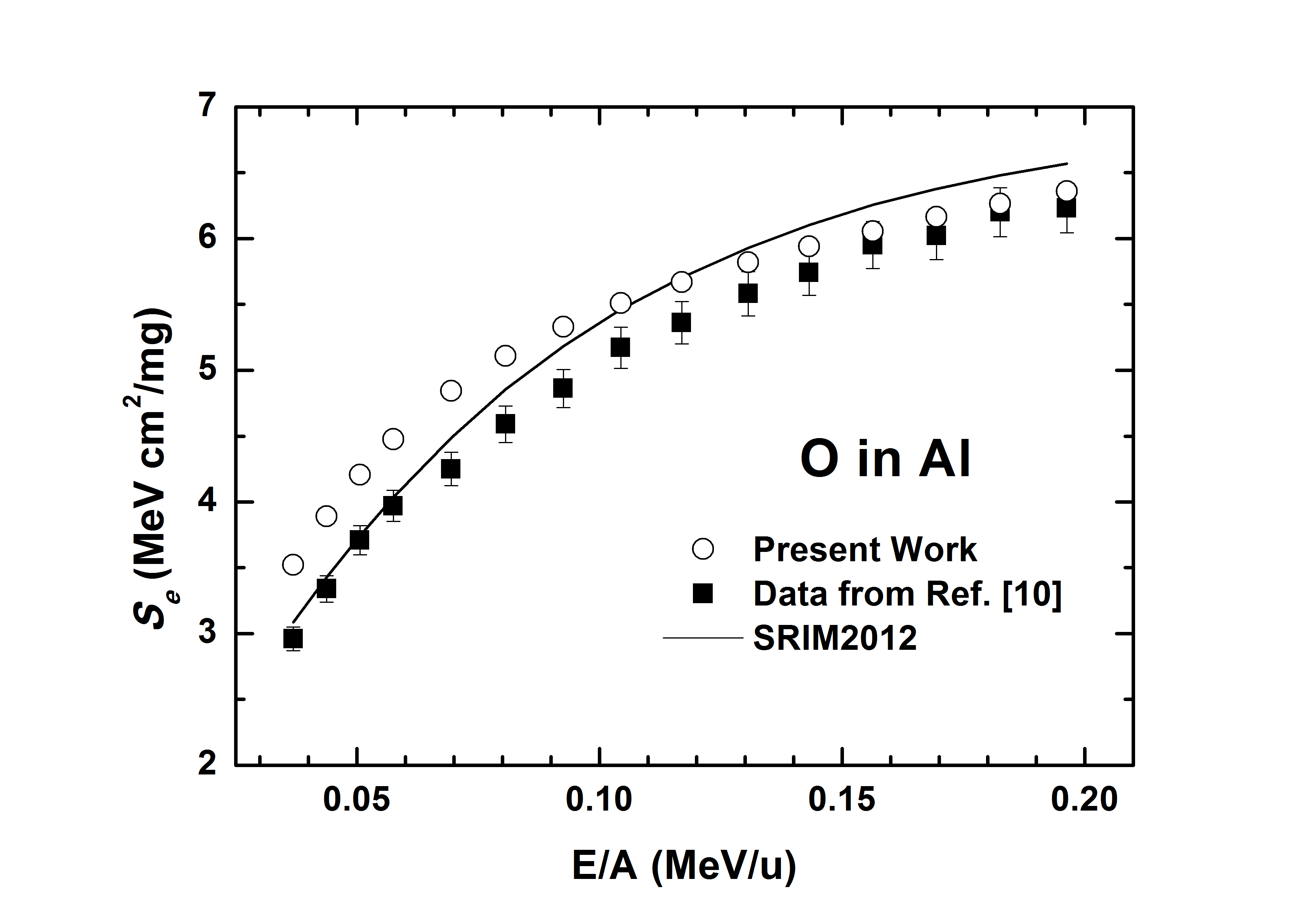}
\caption{Electronic stopping powers of $^{16}$O in Al.}
\label{fig3}
\end{figure}

\begin{figure}
\includegraphics[width=10cm]{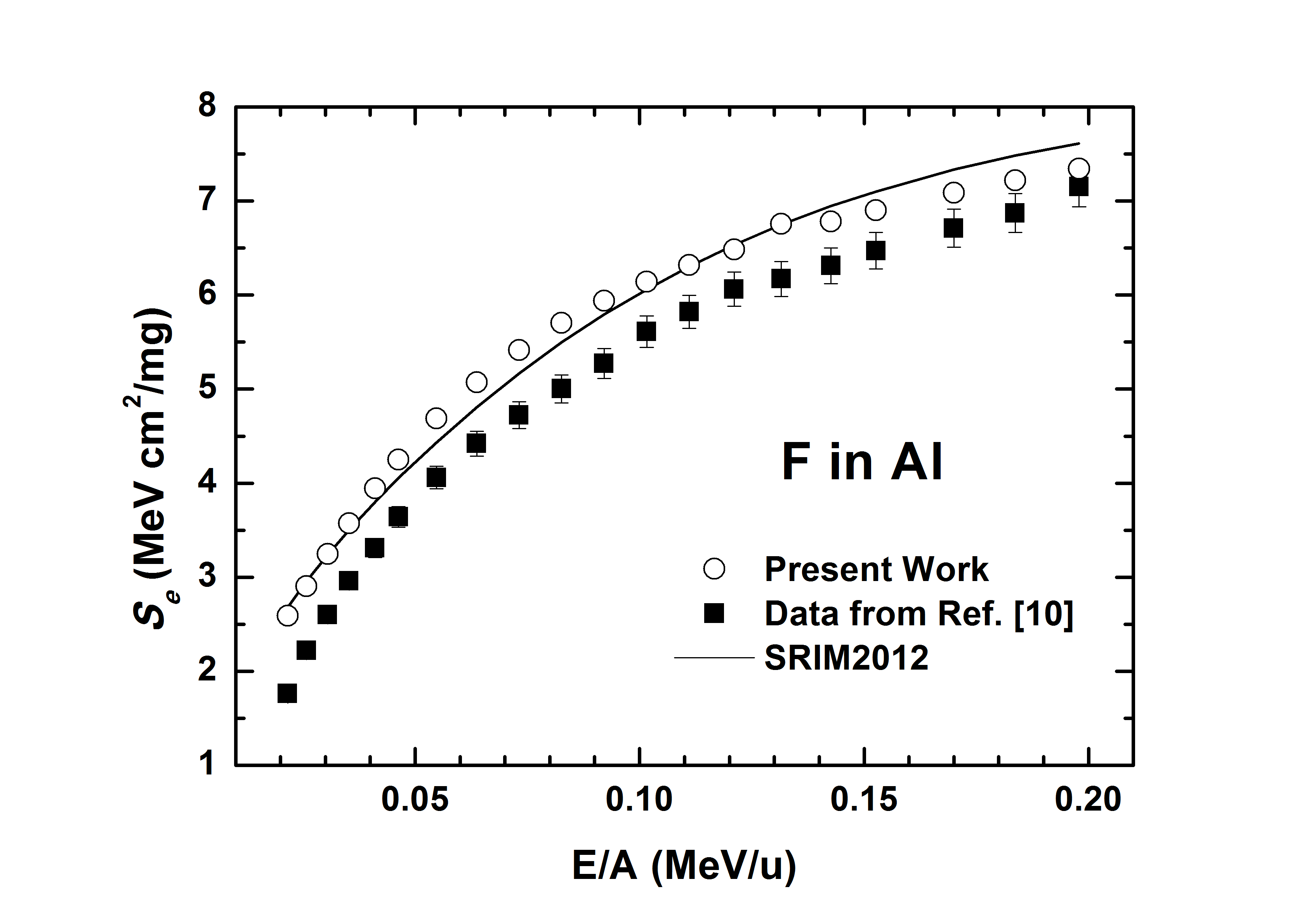}
\caption{Electronic stopping powers of $^{16}$F in Al.}
\label{fig4}
\end{figure}

Electronic stopping powers of $^{16}$O and $^{19}$F ions in Al deduced by Eq. \eqref{eq:one} have been compared with the corresponding experimental data \cite{NIM00Xiting}. The results are shown in Table \ref{table:o16f19} and Figs. \ref{fig3} and \ref{fig4}. The estimated stopping powers of the $^{16}$O ion agree with the corresponding experimental values within less than 5 \% variation above 2 MeV, 6-14 \% between 0.81 and 1.87 MeV (0.05 and 0.12 MeV/u) and 17-19 \% below 0.8 MeV, while the SRIM data agree with them with 5-7 \% variation between 1.11 and 3.14 MeV (0.07 and 0.20 MeV/u) and less than 5 \% below 1 MeV. The estimated stopping powers of the $^{19}$F ion agree with  the corresponding experimental values within 3-9 \% variation above 1.9 MeV, 13-25 \% between 0.58 and 1.75 MeV (0.03 and 0.09 MeV/u), and 31-47 \% below 0.5 MeV, while the SRIM data agree with them with 6-10 \% variation above 1 MeV, 15-33 \% between 0.58 and 0.88 MeV (0.03 and 0.05 MeV/u) and 33-52 \% below 0.5 MeV. The estimation works even better for the $^{16}$O and $^{19}$F ions above 0.12 MeV/u than the SRIM data.

\begin{table}[ht]
\caption{Electronic stopping powers of $^{16}$O and $^{19}$F ions in Al} 
\centering 
\begin{tabular}{l c c c c l c c c} 
\hline\hline 
$^{16}$O &&&&& $^{19}$F \\ [0.5ex]
\hline
E  ~~~~~~~~~~& $\it{S_e}$ & $\it{S_e ^a}$ & $\it{S_e ^b}$  & ~~~~~~~& E~~~~~~~~~~  & $\it{S_e}$ & $\it{S_e ^a}$ & $\it{S_e ^b}$  \\ 
MeV    &      &               MeV/(mg/cm$^2$) & && MeV    &      &              MeV/(mg/cm$^2$) &  \\ [0.5ex] 
\hline
0.59	&3.52 	&2.96	 &3.09 	&&0.41	 &2.59 	&1.76	 &2.68 \\ 
0.70	&3.89 	&3.34	 &3.43 	&&0.49	 &2.90 	&2.22	 &2.96 \\ 
0.81	&4.21 	&3.71	 &3.74 	&&0.58	 &3.25 	&2.60	 &3.24 \\ 
0.92	&4.48 	&3.97	 &4.04 	&&0.67	 &3.57 	&2.96	 &3.50 \\
1.11	&4.84 	&4.25	 &4.49 	&&0.78	 &3.94 	&3.31	 &3.80 \\
1.29	&5.11 	&4.59	 &4.85 	&&0.88	 &4.25 	&3.64	 &4.05 \\
1.48	&5.33 	&4.86	 &5.18 	&&1.04	 &4.69 	&4.06	 &4.43 \\
1.67	&5.51 	&5.17	 &5.46 	&&1.21	 &5.07 	&4.42	 &4.81 \\
1.87	&5.67 	&5.36	 &5.71 	&&1.39	 &5.41 	&4.72	 &5.17 \\
2.09	&5.82 	&5.58	 &5.93 	&&1.57	 &5.70 	&5.00	 &5.49 \\
2.29	&5.94 	&5.74	 &6.10 	&&1.75	 &5.94 	&5.27	 &5.79 \\
2.50	&6.06 	&5.95	 &6.26 	&&1.93	 &6.14 	&5.61	 &6.06 \\
2.71	&6.16 	&6.02	 &6.38 	&&2.11	 &6.32 	&5.82	 &6.30 \\
2.92	&6.26 	&6.20	 &6.48 	&&2.30	 &6.48 	&6.06	 &6.53 \\
3.14	&6.36 	&6.23	 &6.57 	&&2.50	 &6.76 	&6.17	 &6.75 \\
			& &	             && &2.71	 &6.78 	&6.31	 &6.94 \\
			& &	              &&&2.90	 &6.90 	&6.47	 &7.10 \\
			& &	              &&&3.23	 &7.09 	&6.71	 &7.33 \\
			& &	              &&&3.49	 &7.22 	&6.87	 &7.48 \\
			& &	              &&&3.76	 &7.34 	&7.15	 &7.61 \\ [1ex] 
\hline
\end{tabular}
\\{Values of $\it{S_e ^a}$ were taken from Ref. \cite{NIM00Xiting} and those of $\it{S_e ^b}$ obtained from SRIM \cite{SRIM}.}
\label{table:o16f19} 
\end{table}

\begin{figure}
\includegraphics[width=10cm]{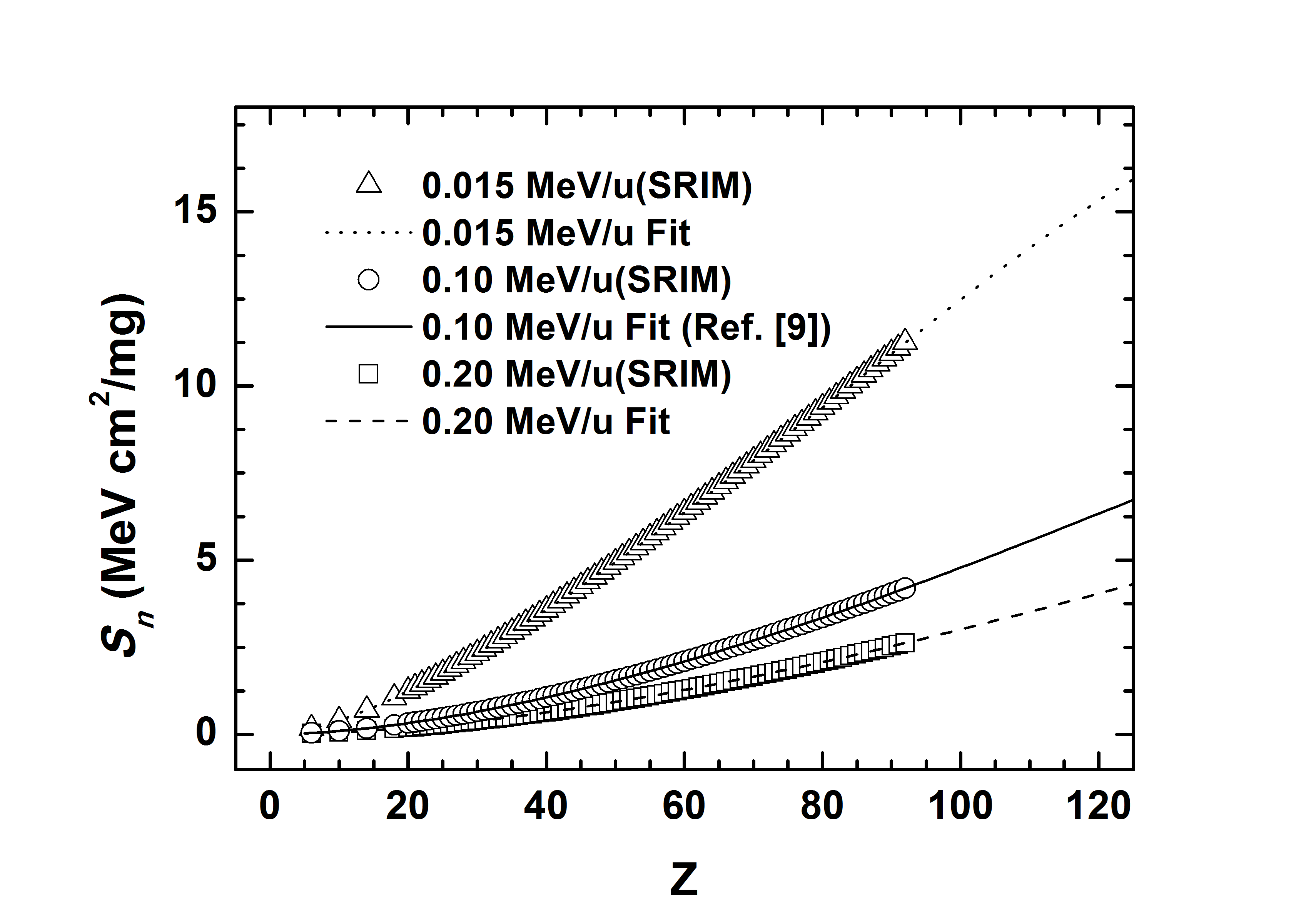}
\caption{Fifth-order polynomial fits for nuclear stopping powers of the ions in Al at 0.015, 0.10 and 0.20 MeV/u. Open symbols are obtained from SRIM \cite{SRIM} and the lines refer to their corresponding fits.}
\label{fig5}
\end{figure}

\begin{figure}
\includegraphics[width=10cm]{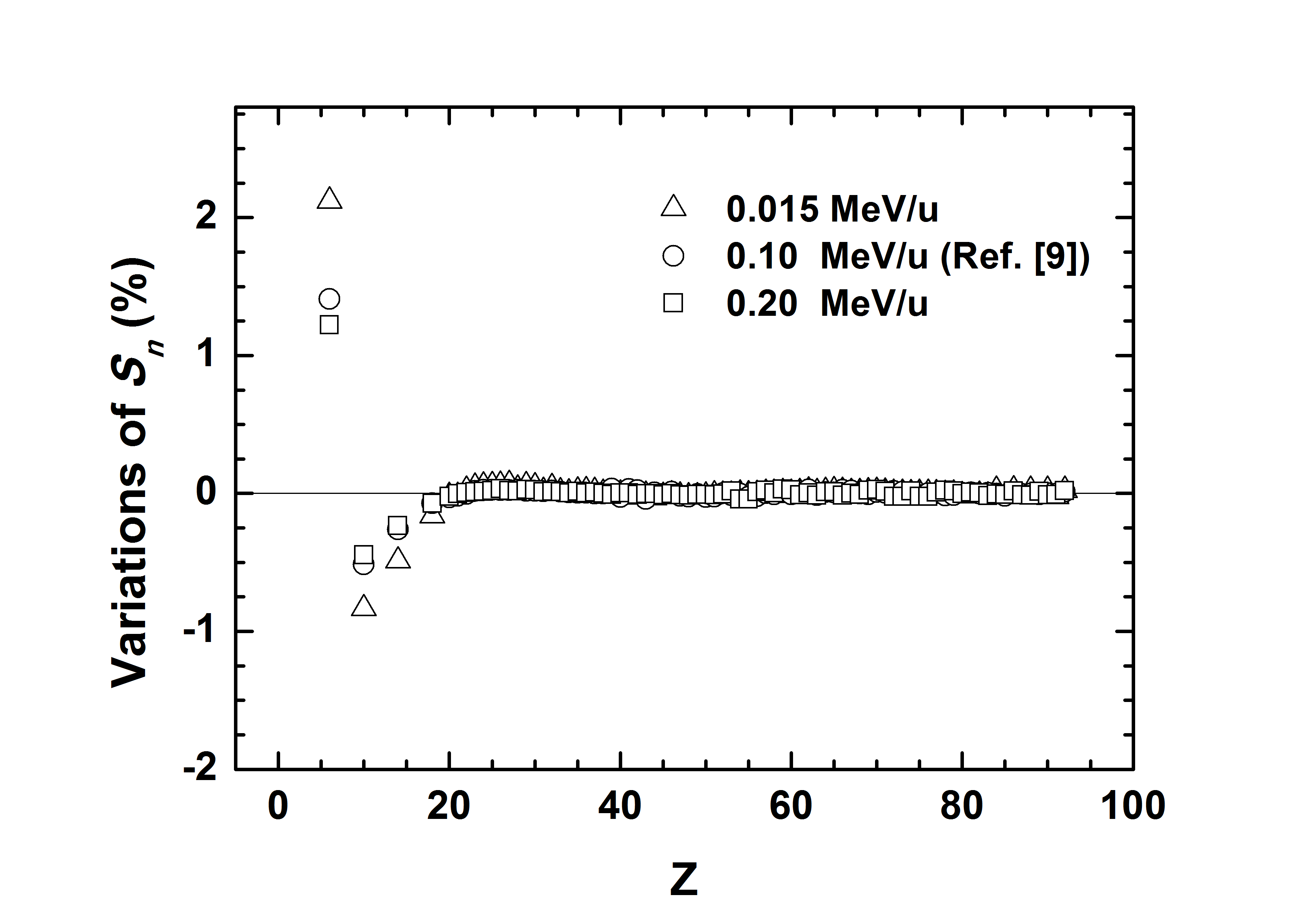}
\caption{Variations of the fits in Fig. \ref{fig5}.}
\label{fig6}
\end{figure}

\begin{figure}
\includegraphics[width=10cm]{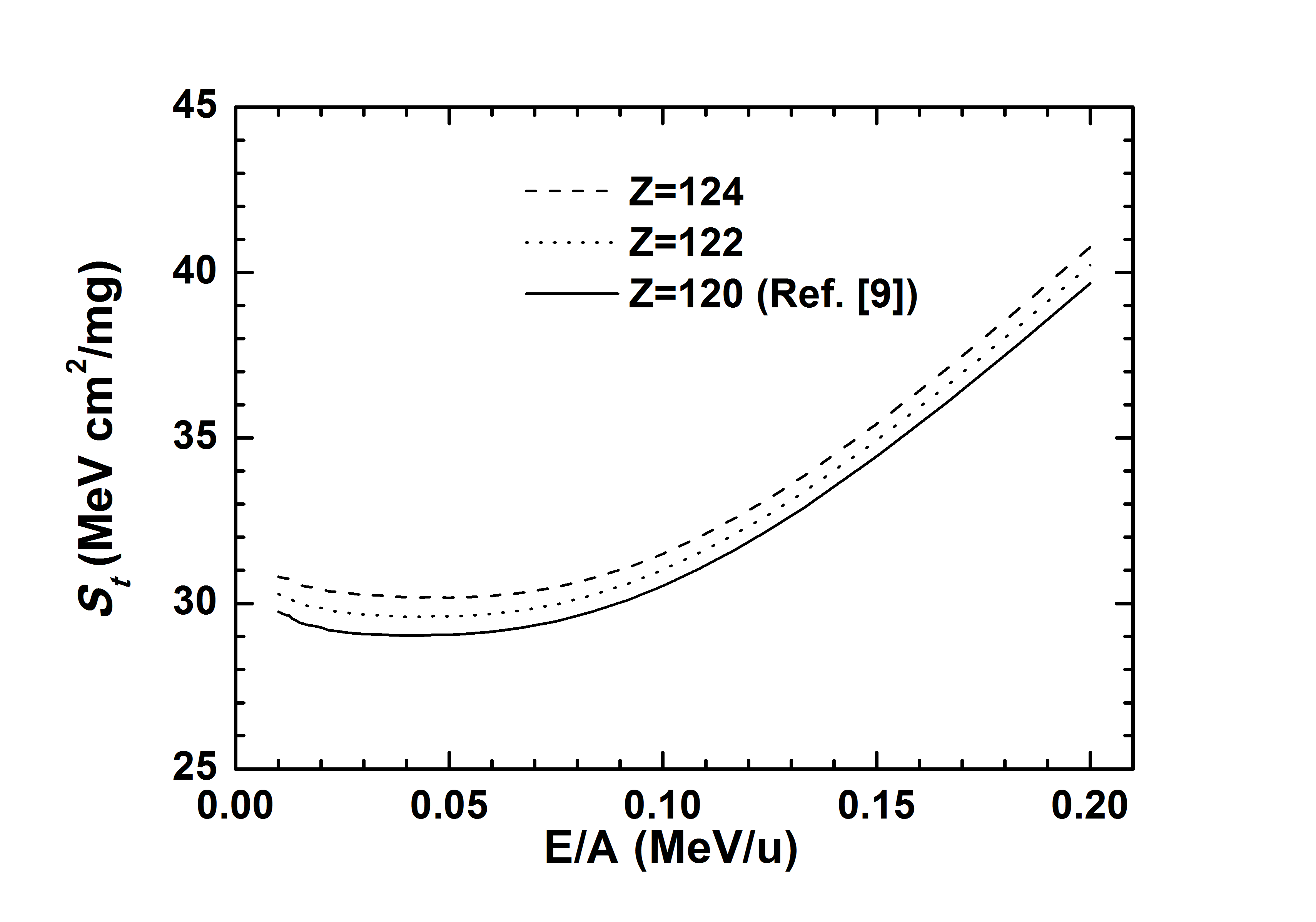}
\caption{Total stopping powers of the superheavy ions with $\it{Z}$=122 and 124 and $\it{A}$=300 in Al at 35 energies ranging from 0.01 to 0.20 MeV/u.}
\label{fig7}
\end{figure}

Nuclear stopping powers of the superheavy ions with $\it{Z}$=122 and 124 and $\it{A}$=300 have been extrapolated by a fifth-order polynomial fit \cite{IJMPE10Chung,CJP14Chung}. The fits and the resulting variations at 0.015, 0.10 and 0.20 MeV/u are shown in Figs. \ref{fig5} and \ref{fig6}, respectively. The fits work well with all the ions within 0.1 \% variation except four light ions, C with 1.2-2.1 \% , Ne with 0.4-0.8 \%, Si with 0.2-0.5 \% and Ar with 0.2 \% variation. The results are listed in Table \ref{table:SheAl} and the total stopping powers are shown in Fig. \ref{fig7}. The electronic and nuclear stopping powers for the $^{300}$122 ion at 0.20 MeV/u are 36.1 and 4.2 MeV/(mg/cm$^2$), respectively, while those at 0.015 MeV/u are 14.4 and 15.6 MeV/(mg/cm$^2$). The corresponding stopping powers of the $^{300}$124 ion at 0.20 MeV/u are 36.5 and 4.3 MeV/(mg/cm$^2$), while those at 0.015 MeV/u are 14.7 and 15.8 MeV/(mg/cm$^2$). The electronic stopping powers are smaller than their corresponding nuclear ones at $\it{E}$$\geq$0.017 MeV/u. Fig. \ref{fig7} shows that total stopping powers of the $^{300}$122 and $^{300}$124 ions tend to decrease slightly until 0.043 MeV/u, that is a little higher than 0.04 MeV/u in the $^{300}$120 ion \cite{CJP14Chung}, and then increase smoothly as the energy increases.

\begin{table}[ht]
\caption{Stopping powers of superheavy ions with $\it{Z}$=120, 122, and 124 and $\it{A}$=300 in Al}  
\centering 
\begin{tabular}{l l l l l l l } 
\hline\hline 
&~~~$\it{Z}$=120\footnote{The data are taken from Ref. \cite{CJP14Chung}. } & &~~~$\it{Z}$=122&&~~~$\it{Z}$=124  \\ [-0.5ex]
\hline 
E ~~~&~~~ $\it{S_e}$ & $\it{S_n}$ ~~~~~&~~~  $\it{S_e}$ & $\it{S_n}$~~~~~ & ~~~$\it{S_e}$ & $\it{S_n}$  \\   [-1.5ex] 
\multicolumn{7}{l}{MeV/u~~~~~~~~~~~~~~~~~~~~~MeV/(mg/cm$^2$)} \\   [-0.5ex]
\hline 
0.010 	&~~~12.745 	&17.002 	&~~~13.045 	&17.237 	&~~~13.348 	&17.457 \\ [-1ex]
0.011 	&~~~12.987 	&16.714 	&~~~13.290 	&16.955 	&~~~13.597 	&17.183 \\ [-1ex]
0.012 	&~~~13.221 	&16.436 	&~~~13.527 	&16.683 	&~~~13.837 	&16.918 \\ [-1ex]
0.013 	&~~~13.448 	&16.182 	&~~~13.758 	&16.437 	&~~~14.070 	&16.681 \\ [-1ex]
0.013 	&~~~13.669 	&15.872 	&~~~13.981 	&16.123 	&~~~14.296 	&16.362 \\ [-1ex]
0.015 	&~~~14.093 	&15.326 	&~~~14.411 	&15.575 	&~~~14.733 	&15.813 \\ [-1ex]
0.017 	&~~~14.498 	&14.858 	&~~~14.822 	&15.112 	&~~~15.149 	&15.357 \\ [-1ex]
0.018 	&~~~14.886 	&14.433 	&~~~15.214 	&14.692 	&~~~15.546 	&14.943 \\ [-1ex]
0.020 	&~~~15.253 	&14.019 	&~~~15.586 	&14.279 	&~~~15.922 	&14.533 \\ [-1ex]
0.022 	&~~~15.584 	&13.605 	&~~~15.920 	&13.861 	&~~~16.259 	&14.111 \\ [-1ex]
0.023 	&~~~15.942 	&13.225 	&~~~16.283 	&13.479 	&~~~16.627 	&13.727 \\ [-1ex]
0.027 	&~~~16.564 	&12.547 	&~~~16.910 	&12.797 	&~~~17.260 	&13.042 \\ [-1ex]
0.030 	&~~~17.130 	&11.940 	&~~~17.481 	&12.184 	&~~~17.835 	&12.424 \\ [-1ex]
0.033 	&~~~17.647 	&11.411 	&~~~18.000 	&11.651 	&~~~18.356 	&11.889 \\ [-1ex]
0.037 	&~~~18.119 	&10.914 	&~~~18.474 	&11.147 	&~~~18.832 	&11.377 \\ [-1ex]
0.040 	&~~~18.552 	&10.471 	&~~~18.907 	&10.698 	&~~~19.265 	&10.922 \\ [-1ex]
0.043 	&~~~18.956 	&10.070 	&~~~19.310 	&10.291 	&~~~19.668 	&10.510 \\ [-1ex]
0.047 	&~~~19.331 	&9.719 	&~~~19.685 	&9.938 	&~~~20.041 	&10.155 \\ [-1ex]
0.050 	&~~~19.682 	&9.365 	&~~~20.033 	&9.575 	&~~~20.387 	&9.784 \\ [-1ex]
0.053 	&~~~20.015 	&9.054 	&~~~20.364 	&9.260 	&~~~20.716 	&9.464 \\ [-1ex]
0.057 	&~~~20.334 	&8.771 	&~~~20.681 	&8.973 	&~~~21.030 	&9.174 \\ [-1ex]
0.060 	&~~~20.640 	&8.505 	&~~~20.984 	&8.702 	&~~~21.331 	&8.899 \\ [-1ex]
0.067 	&~~~21.232 	&8.026 	&~~~21.571 	&8.215 	&~~~21.913 	&8.404 \\ [-1ex]
0.075 	&~~~21.954 	&7.503 	&~~~22.287 	&7.681 	&~~~22.622 	&7.859 \\ [-1ex]
0.083 	&~~~22.677 	&7.072 	&~~~23.006 	&7.245 	&~~~23.336 	&7.418 \\ [-1ex]
0.092 	&~~~23.420 	&6.680 	&~~~23.746 	&6.844 	&~~~24.074 	&7.009 \\ [-1ex]
0.100 	&~~~24.195 	&6.337 	&~~~24.520 	&6.495 	&~~~24.847 	&6.652 \\ [-1ex]
0.108 	&~~~25.002 	&6.038 	&~~~25.328 	&6.190 	&~~~25.656 	&6.342 \\ [-1ex]
0.117 	&~~~25.845 	&5.762 	&~~~26.175 	&5.908 	&~~~26.506 	&6.054 \\ [-1ex]
0.125 	&~~~26.723 	&5.511 	&~~~27.058 	&5.651 	&~~~27.395 	&5.790 \\ [-1ex]
0.133 	&~~~27.633 	&5.294 	&~~~27.975 	&5.429 	&~~~28.319 	&5.565 \\ [-1ex]
0.150 	&~~~29.537 	&4.907 	&~~~29.897 	&5.034 	&~~~30.260 	&5.162 \\ [-1ex]
0.167 	&~~~31.518 	&4.580 	&~~~31.903 	&4.700 	&~~~32.290 	&4.821 \\ [-1ex]
0.183 	&~~~33.561 	&4.295 	&~~~33.974 	&4.409 	&~~~34.390 	&4.523 \\ [-1ex]
0.200 	&~~~35.630 	&4.045 	&~~~36.075 	&4.151 	&~~~36.522 	&4.259 \\ 
\hline 
\end{tabular}
\label{table:SheAl} 
\end{table}

\subsection{Stopping powers of superheavy ions with $\it{Z}$$\geq$120 and $\it{A}$=300 in U}

\begin{figure}
\includegraphics[width=10cm]{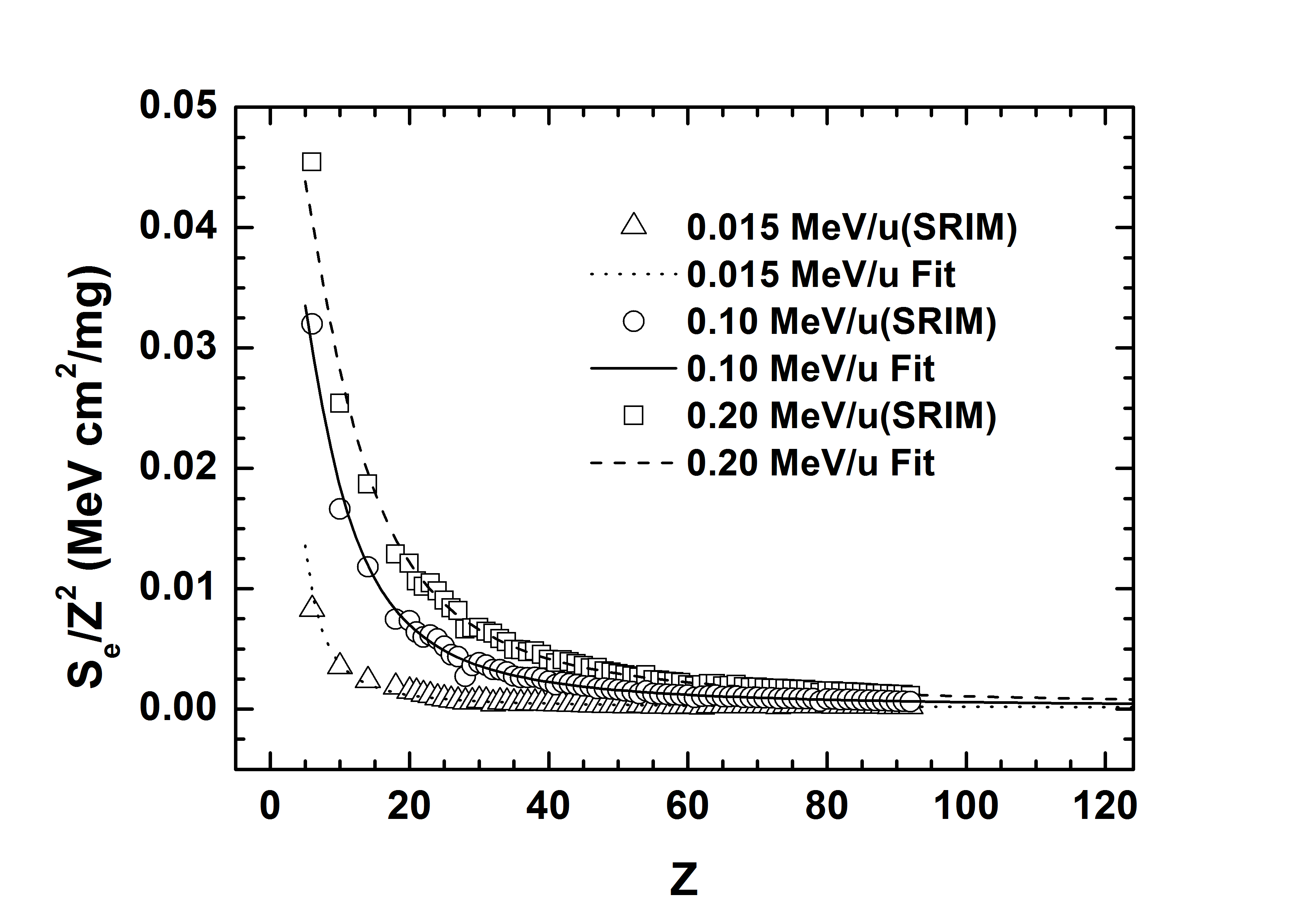}
\caption{Electronic stopping powers of the ions in U divided by $\it{Z^2}$ at 0.015, 0.10 and 0.20 MeV/u. Open symbols are obtained from SRIM \cite{SRIM} and the lines refer to their corresponding fits with Eq. \eqref{eq:one}.}
\label{fig8}
\end{figure}

\begin{figure}
\includegraphics[width=10cm]{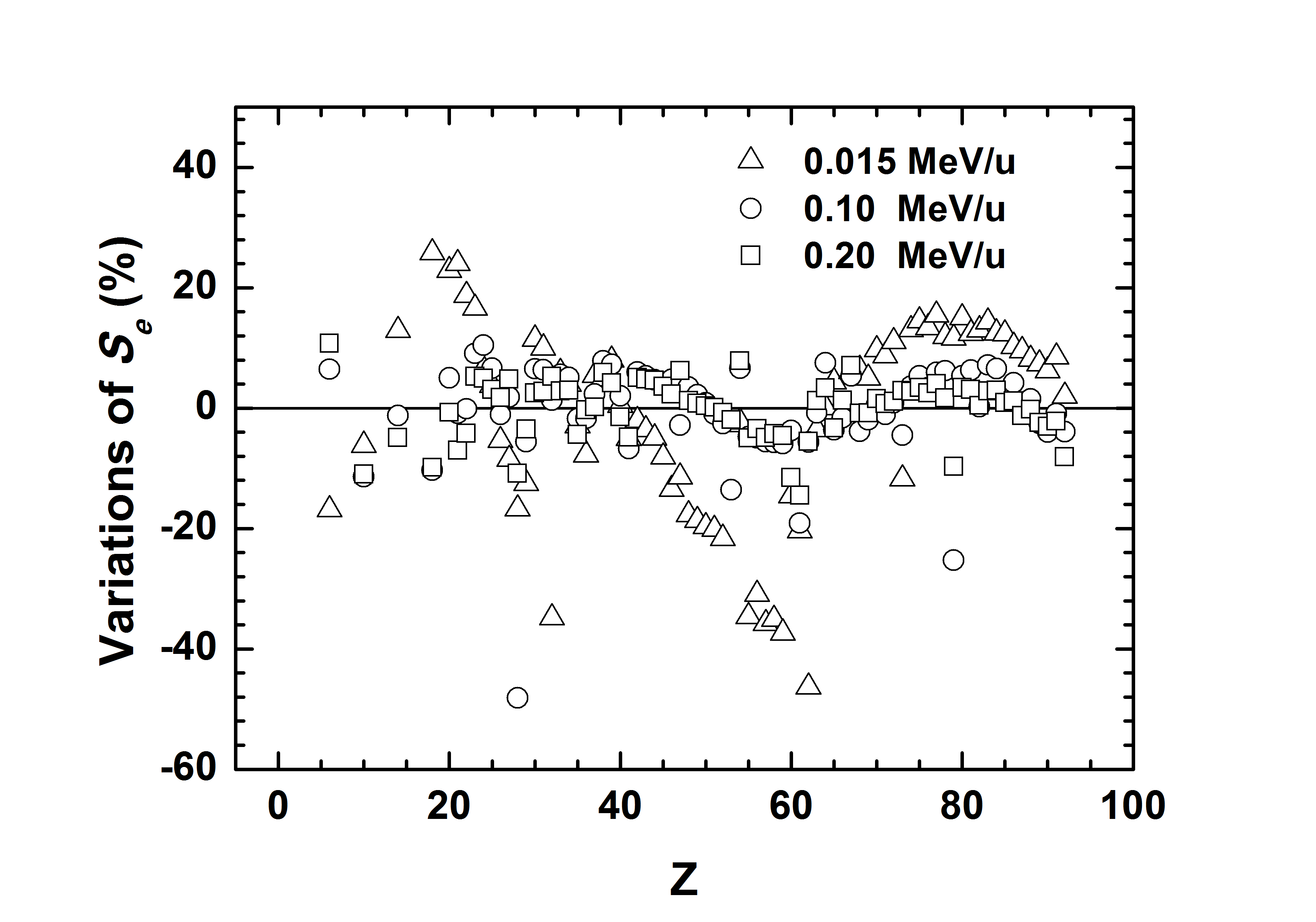}
\caption{Variations of the fits in Fig. \ref{fig8}.}
\label{fig9}
\end{figure}

Electronic and nuclear stopping powers of superheavy ions with $\it{Z}$=120, 122 and 124 and $\it{A}$=300 in U have been estimated similarly as stated in the previous section. The fits for the electronic stopping powers and the resulting variations at 0.015, 0.10 and 0.20 MeV/u are shown in Figs. \ref{fig8} and \ref{fig9}, respectively. They show that the fits work well with most of the ions except six ions with variations larger than 30 \% at  0.015 MeV/u and three ions with variations larger than 14 \% at 0.10 MeV/u. The ions at 0.015 MeV/u are Ge with 35 \%, Cs with 35 \%, Ba with 31 \%, La with 36 \%, Pr with 37 \% and Sm with 46 \% variation, while those at 0.10 MeV/u are Ni with 48 \%, Pm with 19 \% and Au with 25 \% variation. Electronic stopping powers of the Lr ($\it{Z}$=103) ion estimated at 0.015, 0.10 and 0.20 MeV/u are 2.0, 5.9 and 10.9 MeV/(mg/cm$^2$), respectively, while their corresponding values in Northcliffe and Schilling$'$s tables are 1.3, 4.6 and 8.2 MeV/(mg/cm$^2$). Their discrepancies are within a 40 \% variation above 0.04 MeV/u and increase to more than 60 \% variation at the lower energy region.

\begin{figure}
\includegraphics[width=10cm]{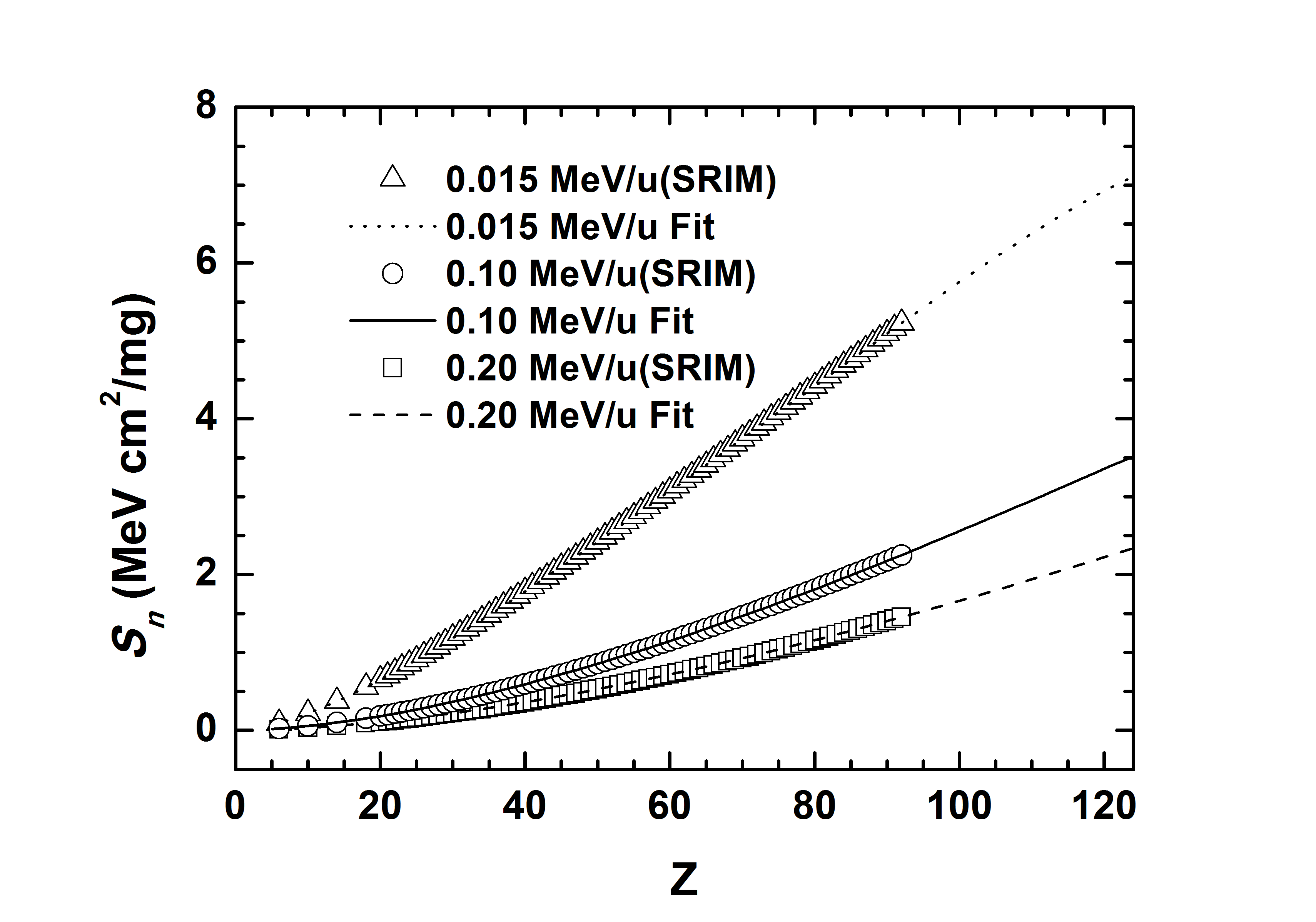}
\caption{Fifth-order polynomial fits for nuclear stopping powers of the ions in Al at 0.015, 0.10 and 0.20 MeV/u. Open symbols are obtained from SRIM \cite{SRIM} and the lines refer to their corresponding fits.}
\label{fig10}
\end{figure}

\begin{figure}
\includegraphics[width=10cm]{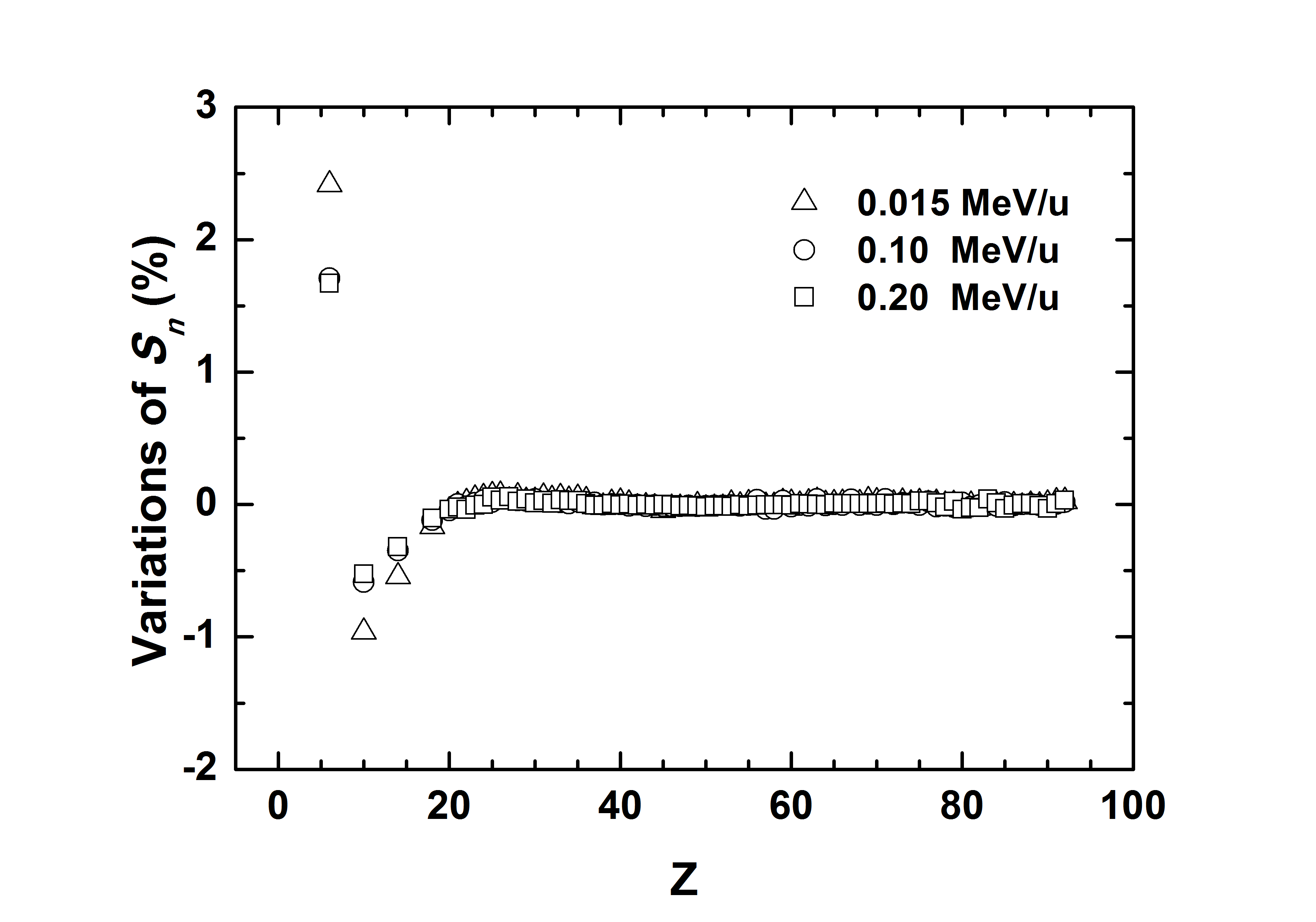}
\caption{Variations of the fits in Fig. \ref{fig10}.}
\label{fig11}
\end{figure}

\begin{figure}
\includegraphics[width=10cm]{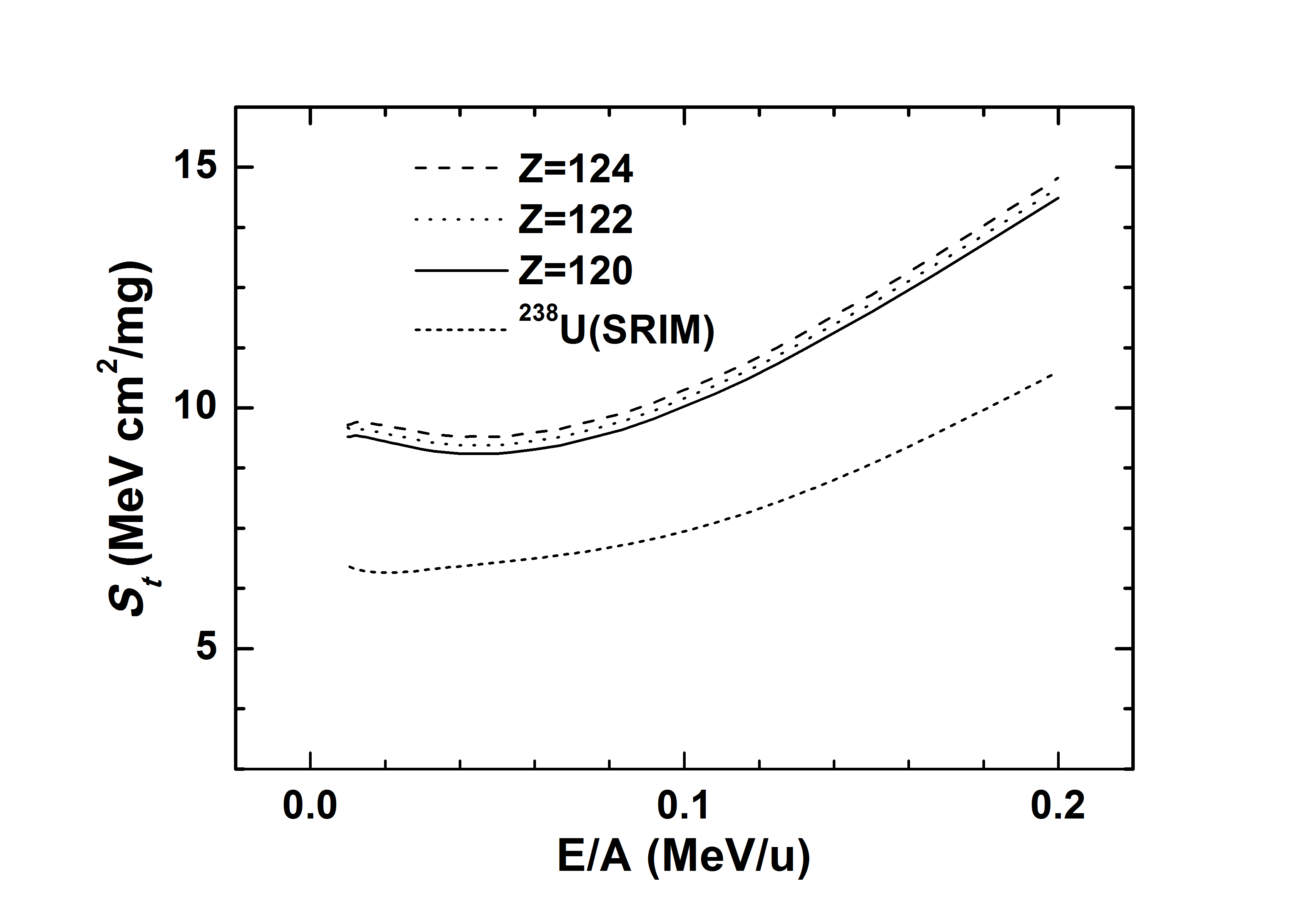}
\caption{Total stopping powers of the superheavy ions with $\it{Z}$=120, 122 and 124 and $\it{A}$=300 in U at  the same energies as in Fig. \ref{fig7}.}
\label{fig12}
\end{figure}

The fits for the nuclear stopping powers and the resulting variations at 0.015, 0.10 and 0.20 MeV/u are shown in Figs. \ref{fig10} and \ref{fig11}, respectively. They work well with all the ions within 0.1 \% variation except four light ions, C with 1.7-2.4 \% , Ne with 0.5-1.0 \%, Si with 0.3-0.5 \% and Ar with 0.1-0.2  \% variation. The results are listed in Table \ref{table:SheU} and the total stopping powers with those of  $^{238}$U are shown in Fig. \ref{fig12}. The electronic and nuclear stopping powers for the $^{300}$120 ion at 0.20 MeV/u are 12.1 and 2.2 MeV/(mg/cm$^2$), respectively, while those at 0.01 MeV/u are 2.0 and 7.4 MeV/(mg/cm$^2$). The corresponding stopping powers of the $^{300}$122 ion at 0.20 MeV/u are 12.3 and 2.3 MeV/(mg/cm$^2$), while those at 0.01 MeV/u are 2.1 and 7.5 MeV/(mg/cm$^2$). The corresponding stopping powers of the $^{300}$124 ion at 0.20 MeV/u are 12.4 and 2.3 MeV/(mg/cm$^2$), while those at 0.01 MeV/u are 2.1 and 7.5 MeV/(mg/cm$^2$). The total stopping powers decrease slightly until about 0.040 MeV/u for $\it{Z}$=120, 0.043 MeV/u for $\it{Z}$=122 and 0.046 MeV/u for $\it{Z}$=124 ion that are higher than 0.02 MeV/u for the $^{238}$U ($\it{Z}$=92) ion and increase smoothly as the energy increases as shown in Al, except a little oscillation below 0.0125 MeV/u. 

\begin{table}[ht]
\caption{Stopping powers of superheavy ions with $\it{Z}$=120, 122, and 124 and $\it{A}$=300 in U} 
\centering 
\begin{tabular}{l l l l l l l } 
\hline\hline 
&~~~$\it{Z}$=120& &~~~$\it{Z}$=122&&~~~$\it{Z}$=124  \\ [-0.5ex]
\hline 
E ~~~&~~~ $\it{S_e}$ & $\it{S_n}$ ~~~~~&~~~  $\it{S_e}$ & $\it{S_n}$~~~~~ & ~~~$\it{S_e}$ & $\it{S_n}$  \\   [-1.5ex]
\multicolumn{7}{l}{MeV/u~~~~~~~~~~~~~~~~~~~~~MeV/(mg/cm$^2$)} \\  [-0.5ex]
\hline
0.010 	&~~~2.041		&7.363	&~~~2.089		&7.512 	&~~~2.138 	&7.512 \\ [-1ex]
0.011 	&~~~2.119		&7.284	&~~~2.168		&7.364 	&~~~2.218 	&7.437 \\ [-1ex]
0.012 	&~~~2.193		&7.231	&~~~2.244		&7.318 	&~~~2.295 	&7.398 \\ [-1ex]
0.013 	&~~~2.265		&7.158	&~~~2.317		&7.248	&~~~2.369 	&7.331 \\ [-1ex]
0.013 	&~~~2.333		&7.076	&~~~2.386		&7.168	&~~~2.440 	&7.252 \\ [-1ex]
0.015 	&~~~2.463		&6.926	&~~~2.519		&7.022 	&~~~2.575 	&7.111 \\ [-1ex]
0.017 	&~~~2.586		&6.776	&~~~2.643		&6.874	&~~~2.702 	&6.966 \\ [-1ex]
0.018 	&~~~2.700		&6.632	&~~~2.760		&6.733	&~~~2.820 	&6.829 \\ [-1ex]
0.020 	&~~~2.810		&6.498	&~~~2.871		&6.602	&~~~2.933 	&6.701 \\ [-1ex]
0.022 	&~~~2.914		&6.359	&~~~2.977		&6.463	&~~~3.041 	&6.563 \\ [-1ex]
0.023 	&~~~3.014		&6.234	&~~~3.079		&6.340	&~~~3.144 	&6.442 \\ [-1ex]
0.027 	&~~~3.205		&5.991	&~~~3.272		&6.098 	&~~~3.339 	&6.202 \\ [-1ex]
0.030 	&~~~3.384		&5.754	&~~~3.453		&5.859 	&~~~3.523 	&5.962 \\ [-1ex]
0.033 	&~~~3.556		&5.543	&~~~3.627		&5.647	&~~~3.699 	&5.748 \\ [-1ex]
0.037 	&~~~3.721		&5.352	&~~~3.794		&5.455 	&~~~3.867 	&5.557 \\ [-1ex]
0.040 	&~~~3.882		&5.168	&~~~3.956 	&5.269	&~~~4.031 	&5.368 \\ [-1ex]
0.043 	&~~~4.032		&5.019	&~~~4.107 	&5.122	&~~~4.183 	&5.223 \\ [-1ex]
0.047 	&~~~4.194		&4.858	&~~~4.270 	&4.958 	&~~~4.348 	&5.057 \\ [-1ex]
0.050 	&~~~4.346		&4.705	&~~~4.424		&4.802 	&~~~4.502 	&4.897 \\ [-1ex]
0.053 	&~~~4.498		&4.575	&~~~4.576 	&4.670 	&~~~4.655 	&4.765 \\ [-1ex]
0.057 	&~~~4.648		&4.459	&~~~4.728		&4.555	&~~~4.808 	&4.649 \\ [-1ex]
0.060 	&~~~4.799		&4.343	&~~~4.879		&4.437	&~~~4.959 	&4.530 \\ [-1ex]
0.067 	&~~~5.100		&4.121	&~~~5.181 	&4.211	&~~~5.263 	&4.299 \\ [-1ex]
0.075 	&~~~5.480		&3.896	&~~~5.563 	&3.985	&~~~5.647 	&4.072 \\ [-1ex]
0.083 	&~~~5.868		&3.678	&~~~5.953 	&3.761 	&~~~6.039 	&3.843 \\ [-1ex]
0.092 	&~~~6.267		&3.502	&~~~6.354 	&3.582 	&~~~6.442 	&3.662 \\ [-1ex]
0.100 	&~~~6.677		&3.356	&~~~6.766 	&3.436 	&~~~6.857 	&3.517 \\ [-1ex]
0.108 	&~~~7.097		&3.202	&~~~7.190		&3.279 	&~~~7.283 	&3.356 \\ [-1ex]
0.117 	&~~~7.529		&3.064	&~~~7.625 	&3.137 	&~~~7.722 	&3.210 \\ [-1ex]
0.125 	&~~~7.971		&2.945	&~~~8.071		&3.016	&~~~8.171 	&3.087 \\ [-1ex]
0.133 	&~~~8.421		&2.847	&~~~8.525 	&2.918	&~~~8.630 	&2.989 \\ [-1ex]
0.150 	&~~~9.339		&2.648	&~~~9.453 	&2.714	&~~~9.568 	&2.780 \\ [-1ex]
0.167 	&~~~10.273	&2.487	&~~~10.399	&2.550	&~~~10.525 	&2.613 \\ [-1ex]
0.183 	&~~~11.21	0	&2.343	&~~~11.348 	&2.403	&~~~11.487	&2.462 \\ [-1ex]
0.200 	&~~~12.140	&2.222	&~~~12.292	&2.279 	&~~~12.444 	&2.337 \\ 
\hline 
\end{tabular}
\label{table:SheU} 
\end{table}

\section{Conclusions}

Electronic and nuclear stopping powers of superheavy ions with $\it{Z}$=122 and 124 and $\it{A}$=300 in Al and $\it{Z}$=120, 122 and 124 and $\it{A}$=300 in U at 35 energies ranging from 0.01 to 0.20 MeV/u have been deduced using the data obtained from SRIM. The total stopping powers of the $^{300}$122 and $^{300}$124 ions in Al tend to decrease slightly until 0.043 MeV/u, that is a little higher than 0.04 MeV/u in the $^{300}$120 ion, and then increase smoothly as the energy increases. The total stopping powers in U decrease slightly until about 0.040 MeV/u for $\it{Z}$=120, 0.043 MeV/u for $\it{Z}$=122 and 0.046 MeV/u for $\it{Z}$=124 ion that are higher than 0.02 MeV/u for the $^{238}$U ion and increase smoothly as the energy increases. Electronic stopping powers of the Lr ion in U estimated at 0.02, 0.10 and 0.20 MeV/u are compared with their corresponding values in Northcliffe and Schilling$'$s tables. Their discrepancies are within a 40\% variation above 0.04 MeV/u and tend to increase at the lower energy region.

The estimated electronic stopping powers of $^{16}$O and $^{19}$F ions agreed well with the experimental values above 0.10 MeV/u, while there were larger deviations near 0.01 MeV/u. This implies that this estimation would be applicable to superheavy ions as well as light ions above 0.10 MeV/u.

\end{document}